\documentclass[a4paper,12pt]{article}
\usepackage[dvips]{graphicx}
\usepackage{subeqn}
\begin{document}
\baselineskip 0.7cm
\def\ie{{\it i.e.}}
\newcommand{\gsim}{ \mathop{}_{\textstyle \sim}^{\textstyle >} } 
\newcommand{\lsim}{ \mathop{}_{\textstyle \sim}^{\textstyle <} } 
\setcounter{footnote}{1}
%%%%%%%%%%%%%%%%%%%%%%%%%%%%%%%%%%%%%%%%%%%%%%
%%%%%%%                                %%%%%%%
%%%%%%%	        Cover page             %%%%%%%
%%%%%%%                                %%%%%%%
%%%%%%%%%%%%%%%%%%%%%%%%%%%%%%%%%%%%%%%%%%%%%%
\thispagestyle{empty}
\vspace*{-15mm}
%----------
\baselineskip 10pt
\begin{flushright}
\begin{tabular}{l}
{\bf OCHA-PP-209}\\
{\bf UT-03-24}\\
{\bf KEK-TH-907}\\
{\bf hep-ph/0308202}
\end{tabular}
\end{flushright}
\baselineskip 18pt 
\vglue 10mm 
%%%%%%%%%%%%%%%%%%%%%%%%%%%%%%%%%%%%%%%%%%%%%%
%                Title 
%%%%%%%%%%%%%%%%%%%%%%%%%%%%%%%%%%%%%%%%%%%%%%
\begin{center}
{\Large\bf
Cosmological gravitino problem confronts 
electroweak physics
}
\vspace{13mm}

\baselineskip 18pt 
\def\thefootnote{\fnsymbol{footnote}}
\setcounter{footnote}{0}
{\bf
Gi-Chol Cho$^{a)}$ and Yosuke Uehara$^{b,c)}$}

\vspace{10mm}

$^{a)}$ {\it Department of Physics, Ochanomizu University, 
Tokyo 112-8610, Japan}
\\
$^{b)}$ {\it Theory Group, KEK, Ibaraki 305-0801 Japan}
\\
$^{c)}$ {\it Department of Physics, University of Tokyo, 
         Tokyo 113-0033, Japan}\\
\vspace{10mm}
\end{center}
%%%%%%%%%%%%%%%%%%%%%%%%%%%%%%%%%%%%%%%%
%%%%%                              %%%%%
%%%%%          Abstract            %%%%%
%%%%%                              %%%%%
%%%%%%%%%%%%%%%%%%%%%%%%%%%%%%%%%%%%%%%%
\begin{center}
{\bf Abstract}\\[7mm]
\begin{minipage}{16cm}

A generic feature of gauge-mediated supersymmetry breaking models 
is that the gravitino is the lightest supersymmetric particle (LSP). 
In order not to overclose the universe, the gravitino LSP should be 
light enough $(\lsim 1~{\rm keV})$, or appropriately heavy 
$(\gsim 1~{\rm GeV})$.  
We study further constraints on the mass of the gravitino imposed by 
electroweak experiments, \ie, muon $g-2$ measurements, electroweak 
precision measurements, and direct searches for supersymmetric 
particles at LEP2. 
We find that the heavy gravitino is strongly disfavored from the lower 
mass bound on the next-to-LSP. 
The sufficiently light gravitino, on the other hand, has rather 
sizable allowed regions in the model parameter space. 
\baselineskip 16pt
\noindent
%%%%%----------------------------------  
\end{minipage}
\end{center}
%%%%%------------------------------------------
%%%%% PACS number(s) & Key words
%%%%%------------------------------------------
%%%\baselineskip 18pt 
%%%{\small 
%%%\begin{flushleft}
%%%{\sl PACS}: 11.30.Pb, 12.15.Lk, 12.60.Jv\\
%%%{\sl Keywords}: 
%%%\end{flushleft}
%%%}
%%%%%---------------------------------- 
\newpage
\renewcommand{\thefootnote}{\arabic{footnote}}
\setcounter{footnote}{0}

Although the standard model (SM) of particle physics is in 
good agreement with the results of high energy collider experiments, 
we expect that new physics beyond the SM lies in the TeV scale, which 
stabilizes the weak scale by protecting the Higgs boson mass from 
radiative corrections. 
The Minimal Supersymmetric extension of the SM (MSSM) is the most 
promising candidate of new physics beyond the SM. 
In the MSSM, quadratic divergences in the radiative corrections 
of the Higgs boson mass are canceled between the contributions from 
the particles in the SM and from their supersymmetric partners. 
However, since exact supersymmetry (SUSY) predicts an unrealistic 
degenerate mass for the ordinary particle and its superpartner, SUSY 
must be broken softly. 
It is, therefore, important to understand the mechanism of SUSY 
breaking and study constraints on the soft SUSY breaking terms from 
phenomenological points of view. 
In particular, serious constraints come from the processes mediated 
by flavor changing neutral current (FCNC), such as 
$K^0$-$\overline{K^0}$ mixing, which require rigorous degeneracy of 
the sfermion masses in the flavor space. 
%%%
%%% a new paragraph

%%%
There are a few classes of SUSY breaking scenarios. 
Among them, gauge mediated SUSY breaking (GMSB) 
models~\cite{GAUGEMEDIATION} have been motivated to satisfy the 
phenomenological constraints on the soft SUSY breaking parameters 
from the FCNC processes. 
In general, the GMSB models consists of 
(i) a secluded sector where supersymmetry is dynamically broken, 
(ii) the visible sector in which all the MSSM fields live, and 
(iii) the messenger fields that transmit the effect of SUSY 
breaking from the secluded sector to the visible sector via the 
ordinary gauge interactions. 
Since the gauge interaction is flavor blind, there is no dangerous 
flavor violating source in the SUSY breaking parameters, and the 
phenomenological constraints from FCNC are satisfied. 
%%%
%%% a new paragraph

%%%
The most striking feature of the GMSB models is in the fact that the gravitino 
is the Lightest Supersymmetric Particle (LSP)\footnote{We assume 
$R$-parity conservation.}. 
In general, the energy density of the stable gravitinos could exceed 
the critical density of the universe, which is so called the 
cosmological gravitino problem~\cite{GRAVITINO1KEV}. 
Since the gravitinos are produced more abundantly as temperature becomes 
higher, the gravitino problem posses a constraint on the reheating 
temperature of the inflation $T_R$. 
The upper bounds on $T_R$ for different mass scales of the gravitino 
mass $m_{3/2}$ are given by 
\cite{GRAVITINOCONSTRAINT,UPPERGRAVITINOMASS} 
%%%
\begin{eqnarray}
T_R \lsim \left\{
  \begin{array}{ll}
   100~{\rm GeV} - 1~{\rm TeV} & \mbox{for $1~{\rm keV} \lsim m_{3/2} 
\lsim 100~{\rm keV}$}
\\[3mm]
\displaystyle{
   10^8~{\rm GeV} \times \left(\frac{m_{3/2}}{1~{\rm GeV}}\right)
\left(\frac{m_{\widetilde{B}}}{100~{\rm GeV}}\right)^{-2}}& 
\mbox{for $m_{3/2} \gsim 100~{\rm keV}$}
  \end{array}
\right. , 
\label{mass}
\end{eqnarray}
%%%
where $m_{\widetilde{B}}$ denotes the bino mass. 
The reheating temperature for the heavier gravitino mass region 
in (\ref{mass}) is compatible with the ordinary inflation scenario 
where it is typically given by $T_R \gsim 10^8~{\rm GeV}$. 
On the other hand, for the lighter gravitino mass region, 
$T_R$ is too low, so that a certain substantial entropy production 
mechanism below $T_R$ should be introduced~\cite{GRAVITINOCONSTRAINT}. 
It should be noted that the overclosure problem due to the gravitino LSP 
is evaded irrelevantly to $T_R$ if the gravitino mass is small enough, 
say, $m_{3/2} \lsim 1~{\rm keV}$~\cite{GRAVITINO1KEV}. 
%%% 
%%% a new paragraph

%%% 
The heavier gravitino LSP ($m_{3/2} \gsim 100~{\rm MeV}$) is also 
imposed a constraint associated with the Next-to-LSP (NLSP). 
The lifetime of NLSP could be comparable with the Big-Bang 
Nucleosynthesis (BBN) era, so that the decay of the NLSP may affect 
the abundance of the light elements in the universe. 
The constraints on $m_{3/2}$ and $T_R$ are examined in 
ref.~\cite{UPPERGRAVITINOMASS} taking into account this effect. 
The allowed regions are given by $m_{3/2} = 5-100~{\rm GeV}$ and 
$T_R=10^9-10^{10}~{\rm GeV}$ when the stau is the NLSP. 
If the neutralino is the NLSP, it gives rise to more severe constraints 
on the reheating temperature because of its small annihilation cross 
section and relatively larger abundance as compared to the stau NLSP. 
%%% 
%%% a new paragraph

%%% 
In this letter, we study constraints on the parameter space of GMSB, 
taking into account the results of muon $g-2$ experiments 
at BNL~\cite{G-2}, the electroweak precision measurements at LEP 
and SLC~\cite{lepewwg}, and direct searches for supersymmetric particles 
at LEP2~\cite{lep_chargino,susy_higgs,sm_higgs}. 
We would like to pay a special attention to whether there are further 
constraints on the gravitino mass scale from those experimental data, 
in addition to the cosmological constraints. 
In the following, we assume that any entropy production mechanisms 
do not exist below $T_R$, so that the cosmologically favored gravitino 
mass scale is limited to $m_{3/2} \lsim 1~{\rm keV}$~\cite{GRAVITINO1KEV} 
or $m_{3/2} = 5 - 100~{\rm GeV}$~\cite{UPPERGRAVITINOMASS}. 
We will show that the allowed region of the gravitino mass is 
sensitive to the muon $g-2$ and the NLSP search experiments. 
The heavy gravitino is allowed only in a small corner of the parameter space. 
%%% 
%%% a new paragraph

%%% 
Let us first briefly review the parameter set of the GMSB models to fix 
our notation. 
The fundamental parameters in the GMSB models can be summarized as 
follows~\cite{GAUGEMEDIATIONREVIEW}:
%%%---------
\begin{eqnarray}
M_m,~ \Lambda,~  k,~  N_m,~ \tan\beta,~ {\rm sgn (\mu)}. 
\label{eq:param_gmsb}
\end{eqnarray}
%%%---------
The first four parameters are related to the SUSY breaking sector and 
the messenger sector. 
$M_m$ is the mass scale of messenger fields and $\Lambda$ denotes the 
scale parameters of soft SUSY breaking terms in the MSSM sector 
at $M_m$, where the messenger fields are integrated out. 
The positivity of the messenger squared mass requires 
$\Lambda<M_m$~\cite{GAUGEMEDIATIONREVIEW}. 
The dimensionless parameter $k(\le 1)$ is the ratio of the fundamental
scale of SUSY breaking and the SUSY breaking scale felt by the messenger 
fields. 
The integer $N_m$ represents the number of messenger fields which transform 
as $\textbf{5}$ + $\bar{\textbf{5}}$ (or $\textbf{10}$ + $\bar{\textbf{10}}$) 
in SU(5), so that the gauge coupling unification is preserved. 
$\tan\beta$ is defined by the ratio of $v_u$ and $v_d$ which are 
the vacuum expectation values of the Higgs fields with the 
hypercharge $Y=1/2$ and $-1/2$, respectively. 
The last parameter in (\ref{eq:param_gmsb}) is the sign of the higgsino 
mass $\mu$. 
The soft SUSY breaking parameters in the MSSM at $M_m$ are expressed in 
terms of the parameters in (\ref{eq:param_gmsb}), and those at the weak 
scale can be obtained by solving the renormalization group equations (RGEs). 
The gravitino mass is given by 
%%%---------
\begin{eqnarray}
m_{3/2} = \frac{\Lambda M_m}{k\sqrt{3} M_{\rm Pl}}, 
\label{gravi_mass}
\end{eqnarray} 
%%%---------
where $M_{\rm Pl}=2.4\times 10^{18}{\rm GeV}$ is the reduced 
Planck mass. 
%%%---------
%%% 
%%% a new paragraph

%%% 
Next we summarize the set of experimental data which we adopt in our 
analysis. 
The anomalous magnetic moment ($g-2$) of the muon has been measured 
precisely at BNL. 
Using the convention $a_{\mu}=(g-2)/2$, the current result is given 
as ~\cite{G-2}
%-----
\begin{eqnarray}
a_{\mu} ({\rm expt})=11659203(8) \times 10^{-10}. 
\end{eqnarray}
%-----
Theoretical prediction on $a_{\mu}$ has a large uncertainty 
due to the hadronic contributions. 
There are a number of estimations on the hadronic contributions using 
various methods. As the SM prediction in our study, we use 
%-----
\begin{eqnarray}
a_{\mu} ({\rm th}) = 11659177(7) \times 10^{-10}. 
\label{eq:g-2_sm}
\end{eqnarray}
%-----
Then the difference between the experimental measurement and the SM 
prediction is given as 
%-----
\begin{eqnarray}
\Delta a_{\mu} =26(10) \times 10^{-10}, 
\label{eq:susy_g-2}
\end{eqnarray}
%-----
which shows 2.6-$\sigma$ discrepancy. 
We require that the SUSY contributions to the muon $g-2$ explain this difference. 
%%%
%%% a new paragraph

%%%
The supersymmetric contributions to the muon $g-2$ come from 1-loop 
diagrams mediated by (i) chargino-sneutrino exchanges and 
(ii) neutralino-smuon exchanges. 
The size of effects from these diagrams is proportional to $\tan\beta$, 
while the sign becomes consistent with (\ref{eq:susy_g-2}) if 
the sign of $\mu$ parameter is positive~\cite{mu-sign}. 
%%%
%%% a new paragraph

%%% 
The electroweak precision measurements of $Z$-pole observables at 
LEP1 and SLC, and the $W$-boson mass at LEP2 and Tevatron, may also 
constrain the parameter space of the GMSB models. 
The electroweak data which we use in our study consist of 17 
$Z$-pole observables and the $W$-boson mass. 
The $Z$-pole observables include 8 line-shape parameters 
($\Gamma_Z, \sigma_h^0, R_\ell, A_{\rm FB}^{0,\ell}(\ell=e,\mu,\tau)$),  
two asymmetries from the $\tau$-polarization data ($A_\tau$, 
$A_e$), the decay rates and the asymmetries of $b$ and 
$c$ quarks ($R_b,R_c,A_{\rm FB}^{0,b},A_{\rm FB}^{0,c}$), 
and the asymmetries measured at SLC $(A_{\rm LR}^0,A_b,A_c)$.  
The experimental data of these observables are summarized in 
ref.~\cite{lepewwg}. 
Taking into account the data for the top-quark mass from 
Tevatron\cite{mtop}, $\alpha_s(m_Z)$~\cite{Hagiwara:fs} 
and $\alpha(m_Z^2)$\cite{BP01}, we find that the SM best fit gives 
$\chi^2/({\rm d.o.f.})=21.4/(21-4)$ (21\% CL). 
The supersymmetric particles affect the electroweak observables 
through the universal gauge-boson propagator corrections 
(oblique corrections) and the process specific vertex or box corrections 
at 1-loop level. 
It has been shown that the contributions from squarks and sleptons 
to the electroweak observables always make the fit to the experimental 
data worse than the SM fit if these particles are as light as 
$\sim 100{\rm GeV}$~\cite{CH2000ew}. 
%%%
%%% a new paragraph 

%%%
In the GMSB models, the NLSP is either the lightest neutralino 
$\widetilde{\chi}^0_1$ or the lighter stau $\widetilde{\tau}_1$. 
As already mentioned, the BBN constraint favors the stau as the NLSP 
rather than the neutralino for the gravitino mass 
$m_{3/2} \gsim 100~{\rm MeV}$~\cite{UPPERGRAVITINOMASS}.  
The lower mass bounds on the NLSP in direct search experiments 
are given as~\cite{ALEPHBOUND}
%%%------------ 
\begin{eqnarray}
m_{\rm NLSP} > \left\{
  \begin{array}{ll}
   55~{\rm GeV} & \mbox{for $\widetilde{\chi}^0_1~{\rm NLSP}$}
\\ [5mm] 
   77~{\rm GeV} & \mbox{for $\widetilde{\tau}_1~{\rm NLSP}$}
  \end{array} 
\right. 
\label{NLSPbound}. 
\end{eqnarray}
%%%--------------------------------------------------
%%%  Fig. 1
%%%--------------------------------------------------
\begin{figure}[t]
\begin{center}
\includegraphics[width=6cm,clip]{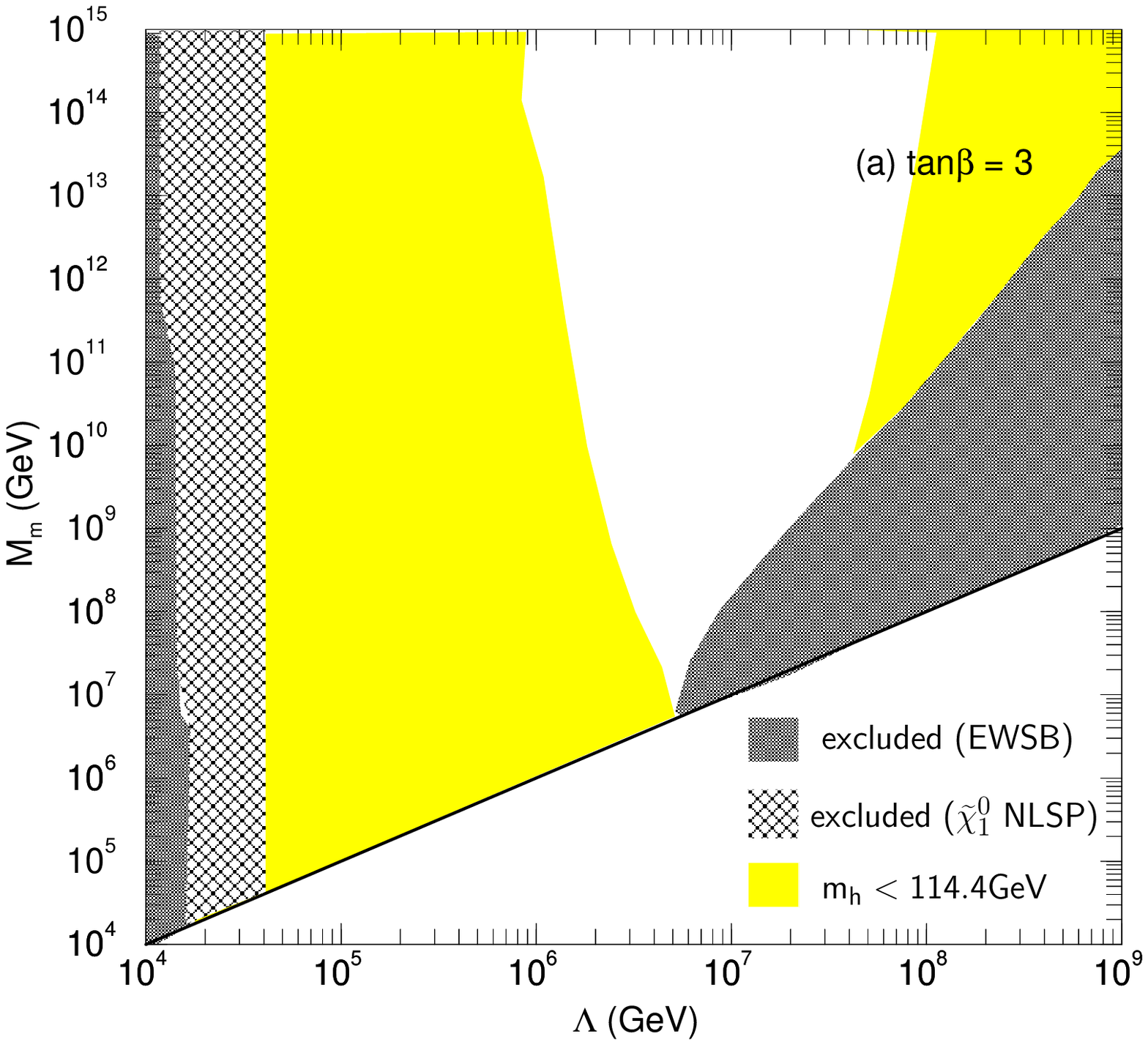}
\includegraphics[width=6cm,clip]{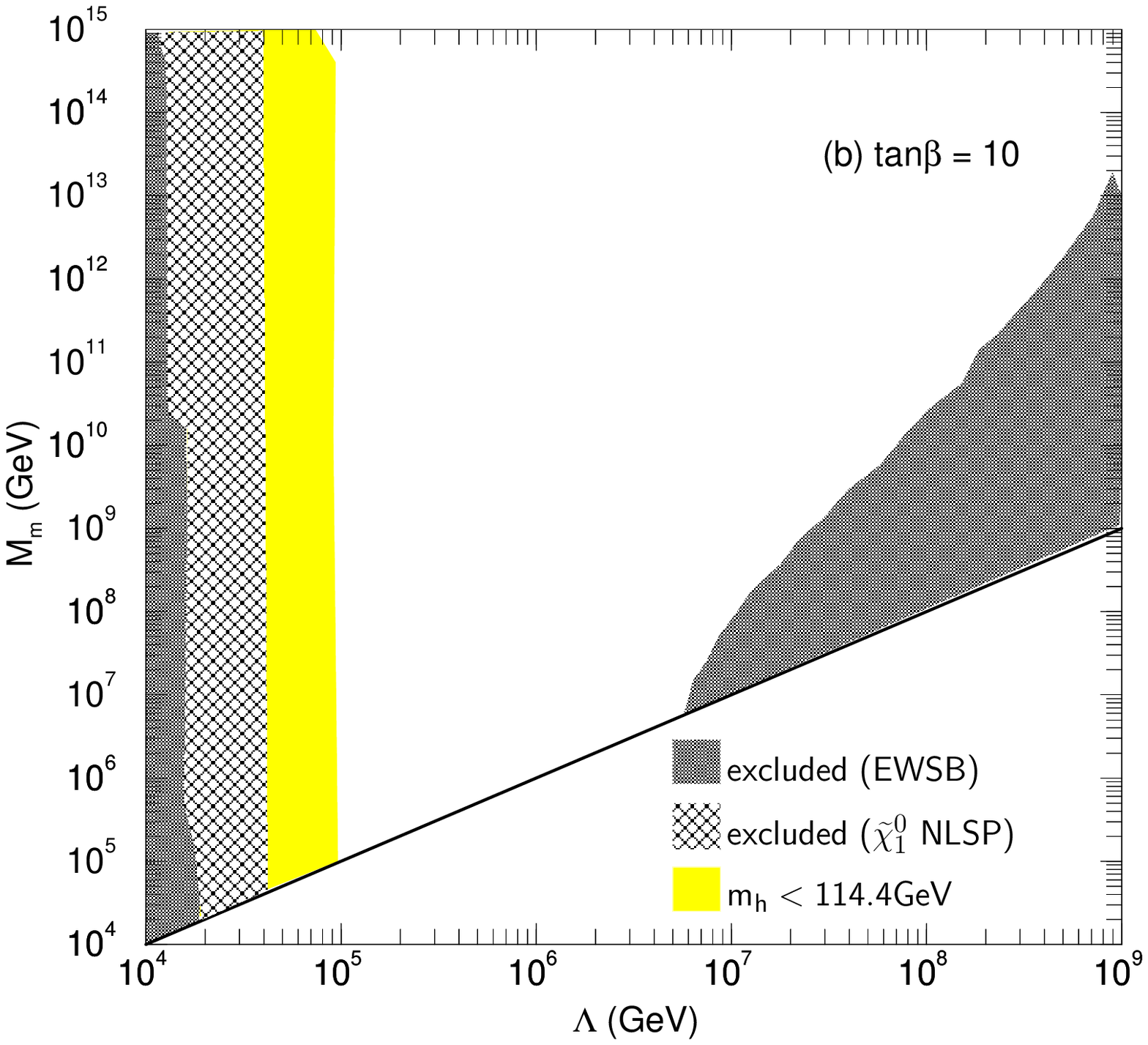}
\includegraphics[width=6cm,clip]{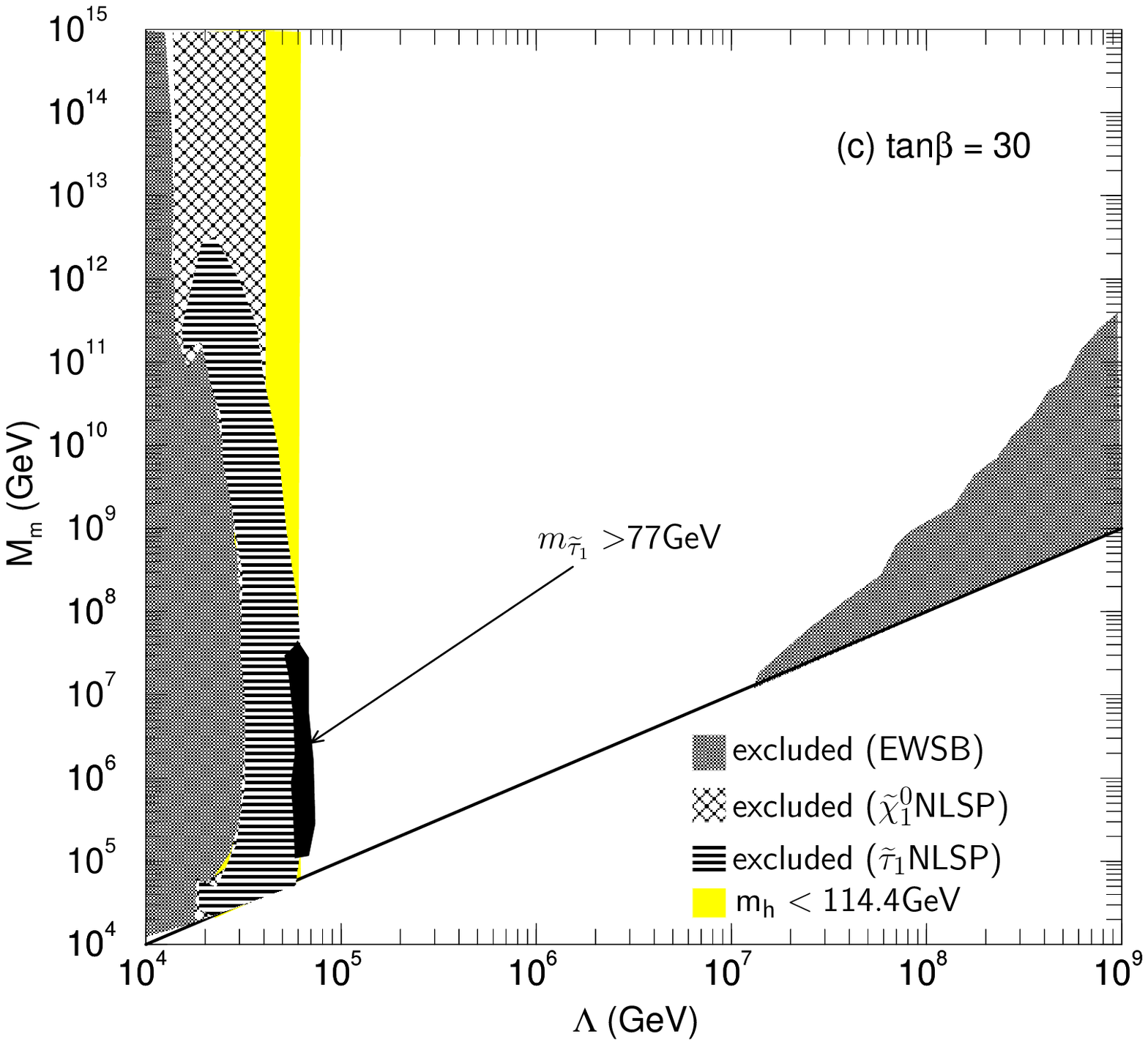}
\includegraphics[width=6cm,clip]{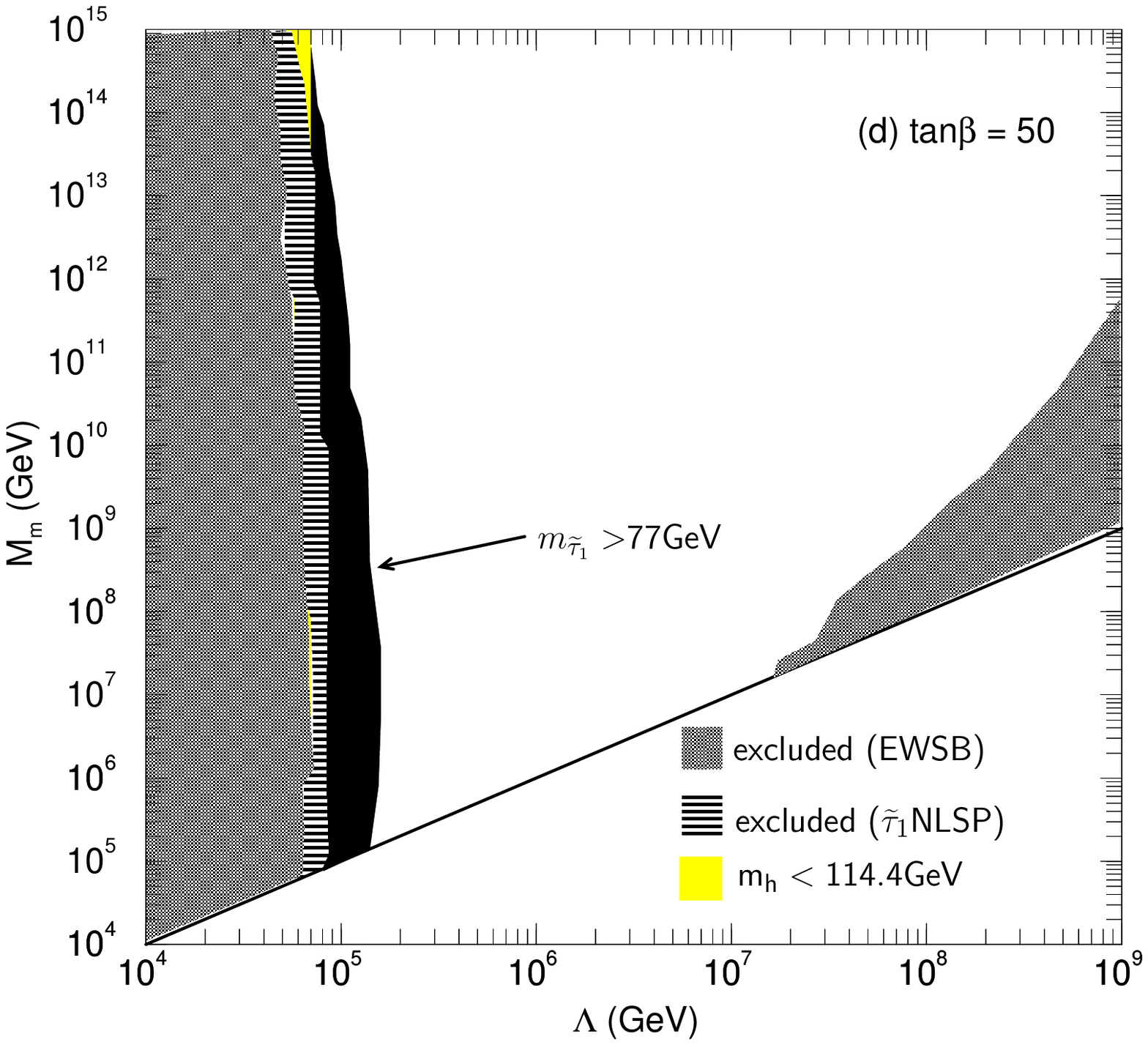}
\end{center}
\caption{
Allowed regions on the $(\Lambda, M_m)$ plane from the radiative 
electroweak symmetry breaking condition and the NLSP direct searches 
for $\tan\beta=3,10,30$ and 50. 
The solid line in each graph shows $M_m = \Lambda$. 
In the dark shaded regions (labeled ``EWSB'') electroweak symmetry is 
not broken radiatively. 
The excluded regions from the direct search limits on the NLSP 
($\widetilde{\chi}^0_1$ or $\widetilde{\tau}_1$) are shown explicitly. 
The light shaded region corresponds to $m_h < 114.4{\rm GeV}$. 
In the blank region above the $M_m=\Lambda$ line, the NLSP is 
the neutralino.  
There are the allowed regions from the direct search limit on the stau 
NLSP~\cite{ALEPHBOUND} in (c) and (d). 
}
\label{NLSP}
\end{figure}
%%%--------------------------------------------------
%%%
%%% a new paragraph 

%%%
In Fig.~\ref{NLSP} we show the allowed region on the $(\Lambda,M_m)$ plane 
from the direct search experiments of the NLSP in (\ref{NLSPbound}). 
In our analysis, we assume radiative electroweak symmetry breaking, 
which is induced by the top-quark contributions to the RGEs for 
(mass)$^2$ terms of the Higgs fields~\cite{ewsb}. 
For simplicity, we fix the parameters $k$ and $N_m$ in 
eq.~(\ref{eq:param_gmsb}) by $N_m = k = 1$. 
We also choose $\mu > 0$ so as to be consistent with the muon $g-2$ 
constraint in (\ref{eq:susy_g-2}). 
The $\tan\beta$ dependence is examined by varying its value as 
$\tan\beta=3,10,30$ and $50$. 
In the figure, the solid line expresses $M_m=\Lambda$, and we 
discuss the region of $M_m >\Lambda$~\cite{GAUGEMEDIATIONREVIEW}. 
%%
%% a new paragraph

%%
The conditions for radiative electroweak symmetry breaking 
exclude the dark regions labelled by ``EWSB''. 
The excluded regions from the direct search limits on the 
$\widetilde{\chi}^0_1$- or $\widetilde{\tau}_1$-NLSP in (\ref{NLSPbound}) 
are shown explicitly.  
In the blank region, the direct search bound on the 
$\widetilde{\chi}^0_1$ NLSP in (\ref{NLSPbound}) is satisfied. 
In the analysis the lower mass bounds on the lighter chargino 
$m_{\widetilde{\chi}^-_1}>104{\rm GeV}$~\cite{lep_chargino} 
and the lightest Higgs boson $m_h >91{\rm GeV}$~\cite{susy_higgs} 
from the LEP2 experiments are used as constraints, 
though they do not reduce the allowed regions of $\widetilde{\chi}^0_1$ 
or $\widetilde{\tau}_1$ NLSP in Fig.~\ref{NLSP}. 
It should be noted that the lower limit on the lightest Higgs boson 
mass $m_h >91{\rm GeV}$ is valid in a very limited parameter space of 
the Higgs sector, and, in most of the parameter space, it coincides 
with the lower mass bound on the SM Higgs boson, 
$m_h > 114.4{\rm GeV}$~\cite{sm_higgs}. 
We find that, if $m_h > 114.4{\rm GeV}$ is used as constraint, 
the allowed region is significantly reduced for 
$\tan\beta=3$ (Fig.~\ref{NLSP}(a)). 
It is remarkable that the allowed region of the stau NLSP 
appears only when $\tan\beta$ is rather large (Figs.~\ref{NLSP}(c) 
and (d)). 
Therefore, the heavier gravitino $m_{3/2} = 5-100{\rm GeV}$ for solving 
the cosmological gravitino problem~\cite{UPPERGRAVITINOMASS} is strongly 
constrained from the stau NLSP search experiments. 
%%%------------
%%%
%%% a new paragraph 

%%%
%%%--------------------------------------------------
%%%  Fig. 2
%%%--------------------------------------------------
\begin{figure}[t]
\begin{center}
\includegraphics[width=7cm,clip]{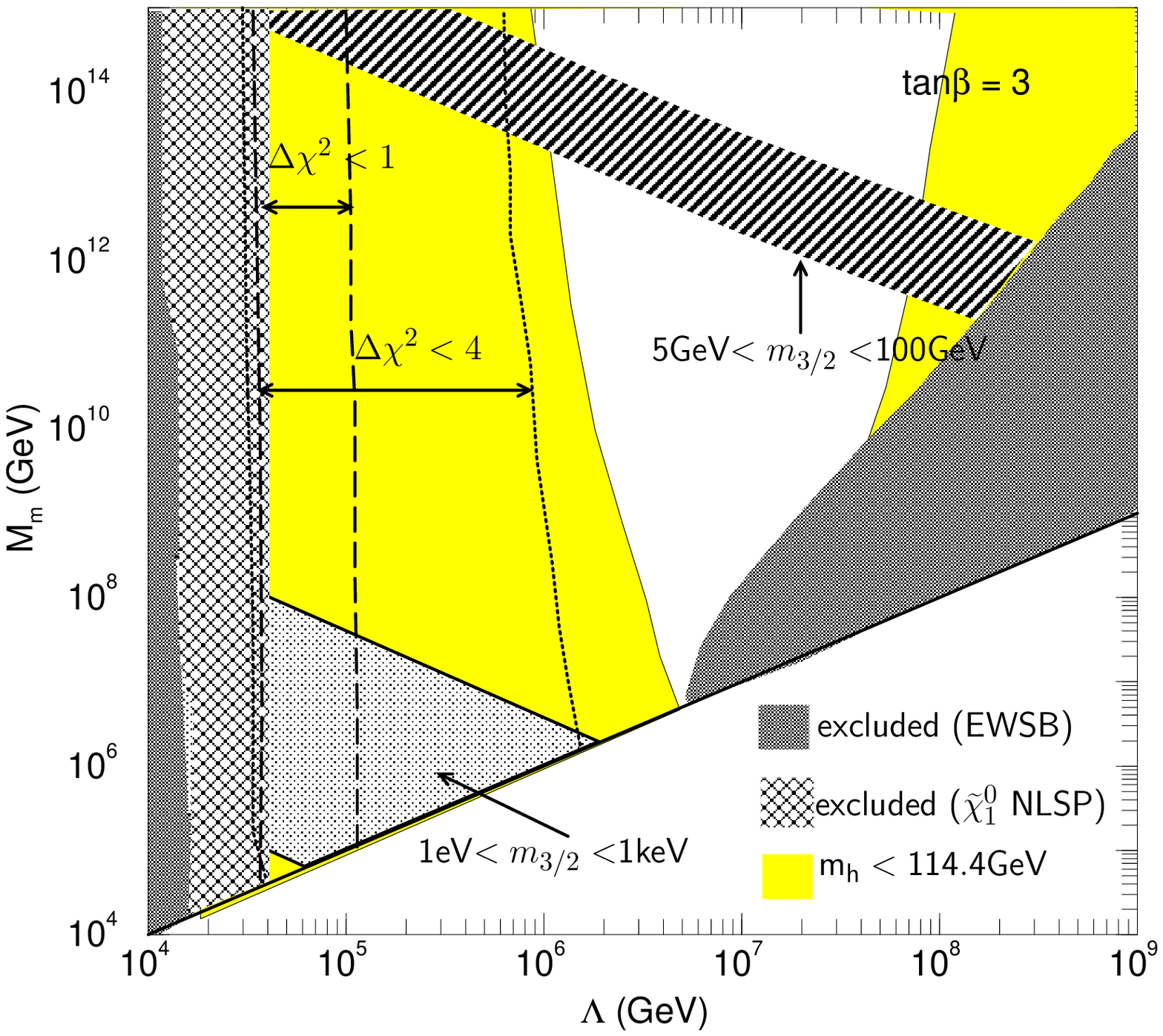}
\includegraphics[width=7cm,clip]{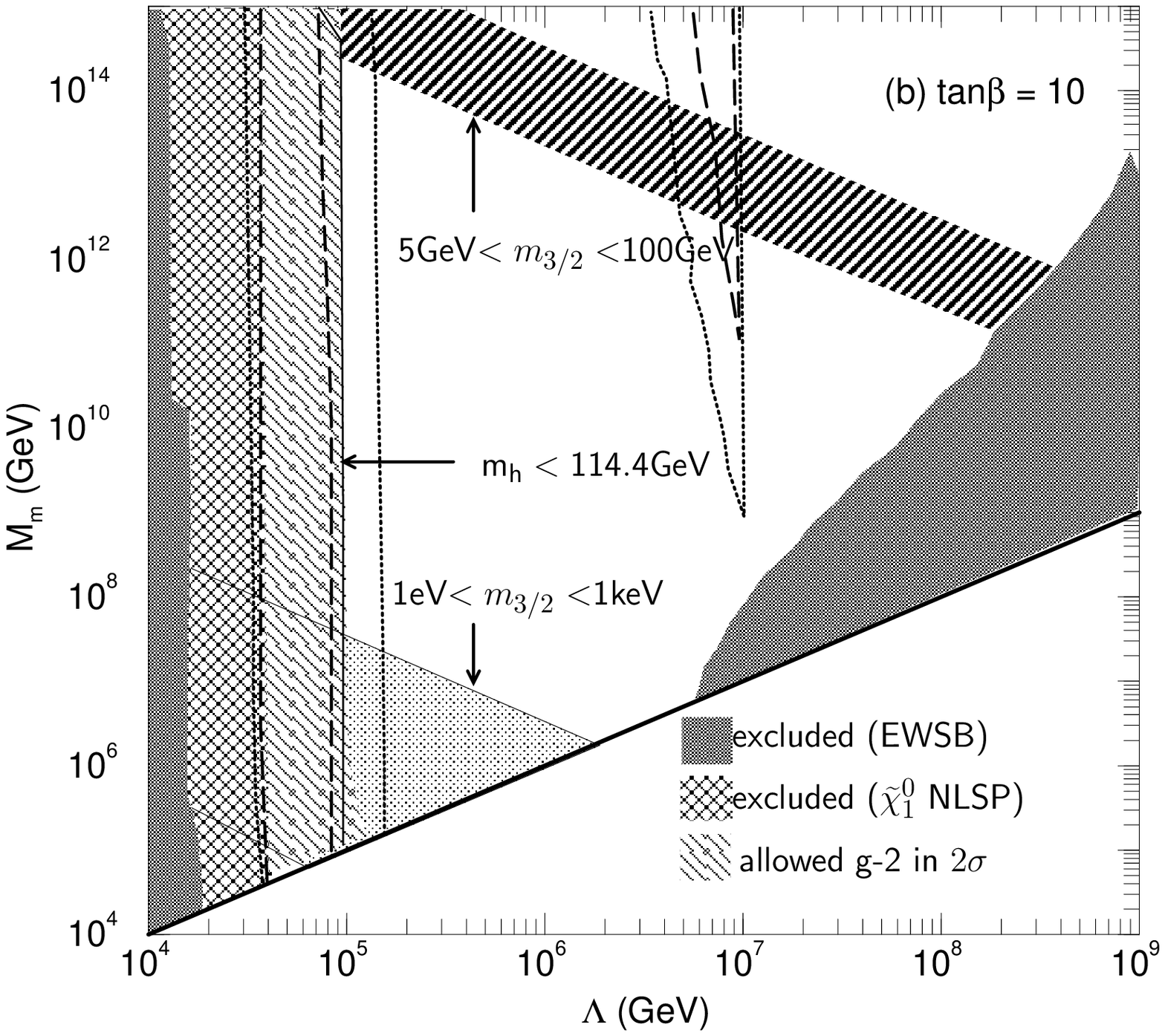}
\end{center}
\caption{
Constraints on the $(\Lambda,M_m)$ plane from the electroweak 
precision measurements and the muon $g-2$ experiments for 
$\tan\beta=3$ (a) and 10 (b). 
The gravitino mass range is shown for $1{\rm eV} < m_{3/2} < 1{\rm keV}$  
and $5 {\rm GeV} < m_{3/2} < 100{\rm GeV}$. 
The enclosed regions by the dotted lines give $\Delta \chi^2 < 4$, 
while those by the long-dashed lines give $\Delta \chi^2 < 1$ for 
the electroweak precision data. 
The 2-$\sigma$ allowed region of the muon $g-2$ experiments is 
shown explicitly in (b). 
In (a), the allowed region of the muon $g-2$ is hidden by the 
$\widetilde{\chi}^0_1$ NLSP excluded region. 
}
\label{FULL-1}
\end{figure}
%%%--------------------------------------------------
Let us examine constraints on the GMSB models from the muon $g-2$ and 
the electroweak precision data for $\tan\beta=3$ and $10$ in 
Fig.~\ref{FULL-1}. 
On the NLSP constraints, we superpose the gravitino mass ranges 
($1~{\rm eV} < m_{3/2} < 1~{\rm keV}$ and 
$5~{\rm GeV} < m_{3/2} < 100~{\rm GeV}$) and the 2-$\sigma$ allowed 
regions of the muon $g-2$ data in the $(\Lambda,M_m)$ plane. 
The long-dashed and dotted lines indicate $\Delta \chi^2 \equiv 
\chi^2_{\rm SUSY}-\chi^2_{\rm SM} =1$ and $4$ in the fit to the 
electroweak precision data, respectively. 
It is easy to see that there is no allowed region of the muon $g-2$ 
data in Fig.~\ref{FULL-1}(a) ($\tan\beta=3$). 
As is already mentioned, the SUSY contributions to the muon $g-2$ 
are proportional to $\tan\beta$. 
If $\tan\beta$ is small, therefore, relatively light SUSY particles 
are required for sizable contributions to the muon $g-2$. 
Such parameter region in Fig.~\ref{FULL-1}(a) is inconsistent with 
the direct search limit on the $\widetilde{\chi}^0_1$ NLSP mass. 
If $\tan\beta$ becomes larger (Fig.~\ref{FULL-1}(b)), 
we find an allowed region for the light gravitino, 
$m_{3/2}<1~{\rm keV}$, where constraints from the muon $g-2$, 
the electroweak precision measurements, and the direct search for 
the $\widetilde{\chi}^0_1$ NLSP are satisfied simultaneously. 
However the allowed region is significantly reduced if the lower mass bound 
on the lightest Higss mass is given by $m_h > 114.4{\rm GeV}$. 
%%%--------------------------------------------------
%%%  Fig. 3
%%%--------------------------------------------------
\begin{figure}[t]
\begin{center}
\includegraphics[width=7cm,clip]{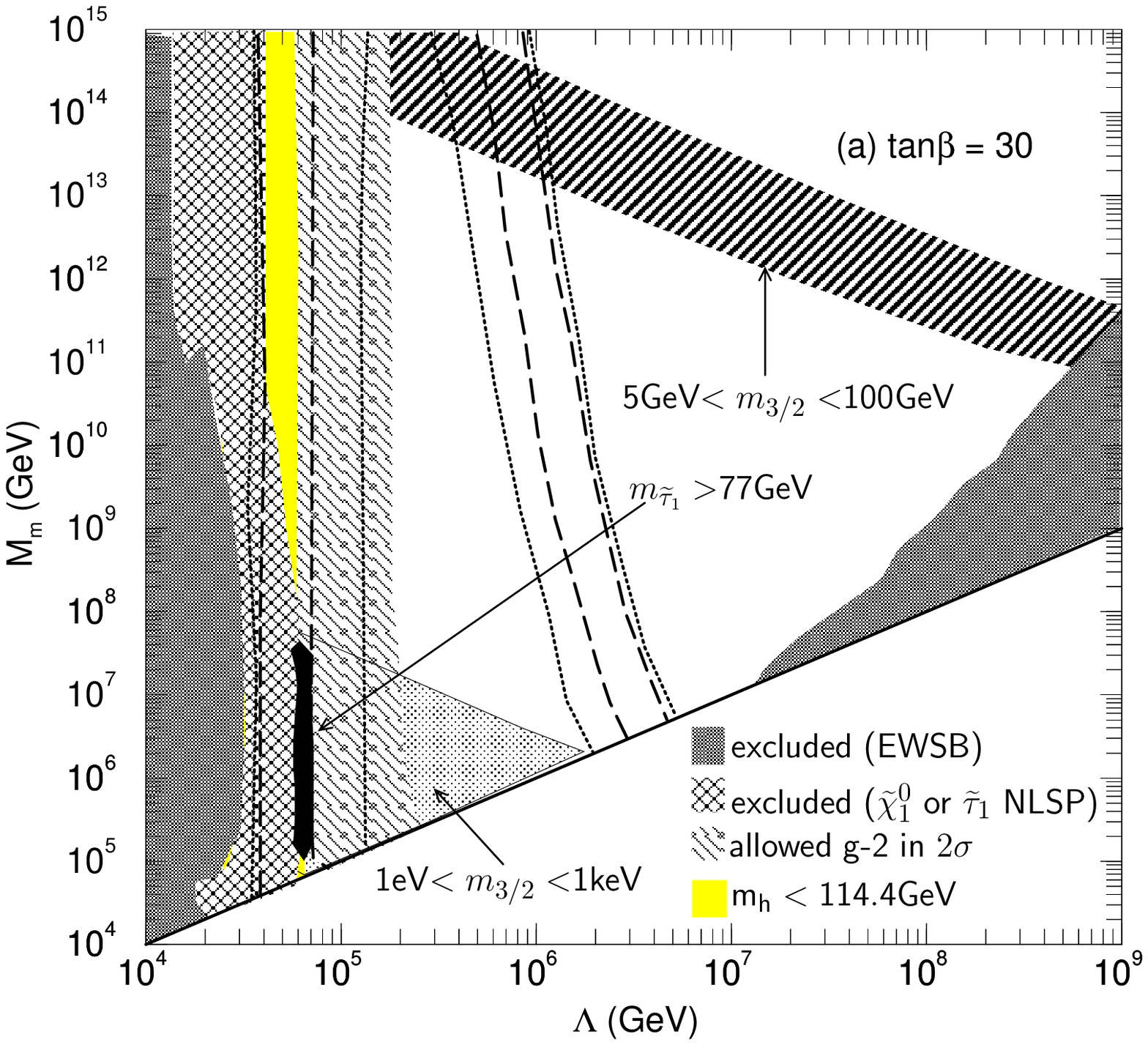}
\includegraphics[width=7cm,clip]{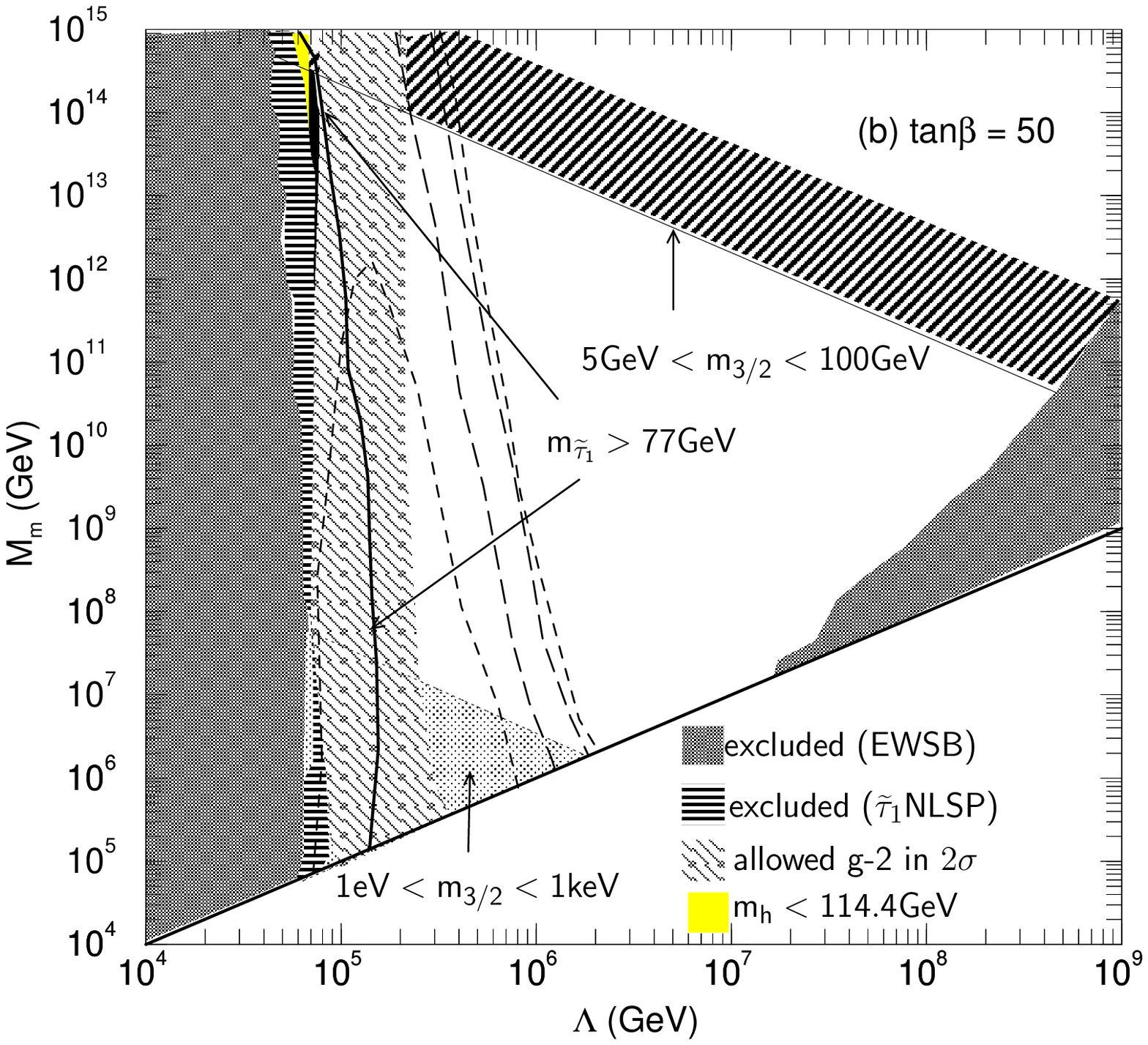}
\end{center}
\caption{
Constraints on the $(\Lambda,M_m)$ plane from the electroweak 
precision measurements and the muon $g-2$ experiments for 
$\tan\beta=30$ (a) and 50 (b). 
The gravitino mass range is shown for $1{\rm eV} < m_{3/2} < 1{\rm keV}$  
and $5 {\rm GeV} < m_{3/2} < 100{\rm GeV}$. 
The enclosed regions by the dotted lines give $\Delta \chi^2 < 4$,  
while those by the long-dashed lines give $\Delta \chi^2 < 1$.  
The 2-$\sigma$ allowed region of the muon $g-2$ experiments is 
shown explicitly. 
In the enclosed region by the thick solid-line in (b), the 
$\widetilde{\tau}_1$ mass satisfies the direct search limit of 
the NLSP, {\it i.e.},  $m_{\widetilde{\tau}_1} > 77{\rm GeV}$. 
}
\label{FULL-2}
\end{figure}
%%%--------------------------------------------------
%%%------------
%%%
%%% a new paragraph 

%%%
In Fig.~\ref{FULL-2}, we show constraints on the model parameter space 
for $\tan\beta=30$ (a) and $50$ (b). 
For $\tan\beta=30$, we find that, in a sizable region, the lighter  
gravitino is consistent with all the experimental constraints. 
The heavier gravitino, however, is again disfavored because of 
the lower mass bound on the stau NLSP from collider experiments. 
Fig.~\ref{FULL-2}(b) shows that the fit to the electroweak precision 
data at the lighter gravitino region could be worse ($\Delta \chi^2 > 4$) 
than the case for smaller $\tan\beta(\le 30)$. 
For the heavier gravitino, there is a very small region which is 
compatible with the bounds from the stau NLSP and the muon $g-2$. 
From these analyses, we find that the lower mass bound on the stau 
NLSP gives the most stringent constraint on the heavier gravitino, 
which could be allowed only for large $\tan\beta$, say, 
$\tan\beta \sim 50$. 
%%%------------
%%%
%%% a new paragraph 

%%%
%%%------------
We have so far performed our analysis by fixing the parameters 
$k$ and $N_m$ in (\ref{eq:param_gmsb}) to be unity. 
It may be helpful to mention the $k$ or $N_m$ dependences of our 
analysis. 
First, the $k$ parameter is related to the gravitino mass through
(\ref{gravi_mass}). 
If $k$ is smaller than 1, the gravitino mass increases for fixed 
values of $\Lambda$ and $M_m$. 
This means that the gravitino mass range on the $(\Lambda, M_m)$ plane 
in our study is lowered for $k<1$, in parallel with the range for 
$k=1$. 
It is easy to see that the constraints on both the heavier and lighter 
gravitinos are not altered so much for $k<1$. 
The dependence on $N_m$ of the result is rather complicated because 
it reflects the detail of the SUSY breaking sector. 
In general, the soft SUSY breaking parameters tend to be large as 
$N_m$ increases, so that the constraints on $(\Lambda,M_m)$, \ie, 
the NLSP mass, may be weaker for $N_m>1$. 
%%%------------
%%%
%%% a new paragraph : summary

%%%
In summary, we have studied constraints on the model parameter space 
of GMSB taking into account the muon $g-2$ experiments, the electroweak 
precision measurements and the direct search experiments on the NLSP.  
The main interest of our study is in the influence of these experimental 
results on the allowed gravitino mass scales which are obtained from the 
cosmological gravitino problem. 
Assuming no entropy production mechanism below the reheating temperature 
of the inflation, we focused on two different gravitino mass scales, 
$m_{3/2}<1{\rm keV}$~\cite{GRAVITINO1KEV} and 
$5{\rm GeV} < m_{3/2}<100{\rm GeV}$~\cite{UPPERGRAVITINOMASS}. 
We find that both possibilities are disfavored from the muon $g-2$ data 
and/or the NLSP direct search experiments for $\tan\beta=3$.  
The light gravitino mass range could be allowed by these 
experiments in the small parameter region for $\tan\beta=10$. 
However it is significantly reduced if $m_h > 114.4~{\rm GeV}$ 
is used as the lower mass bound on the lightest Higgs boson. 
For $\tan\beta>30$, the light gravitino mass is compatible with the 
low-energy experiments in sizable parameter regions. 
On the other hand, the heavier gravitino is strongly disfavored 
by the lower mass bound on the stau NLSP and the muon $g-2$ 
experiments. 
It could be allowed only if $\tan\beta$ is large enough, say, 
$\tan\beta\sim 50$. 
The possibility of the heavier gravitino, therefore, is pushed to a small 
corner of the parameter space. 

%%%------------
%%%
%%% a new paragraph 

%%%
%%%\section*{Acknowledgement}
\vspace{0.5cm}

The authors thank K.~Hamaguchi and M.~Fujii for stimulating discussions, 
and N. Oshimo for careful reading of the manuscript. 
They are also grateful to C. Kao for allowing us to use his computer 
program to solve the RGE of GMSB. 
Y.U. also thanks Japan Society for the Promotion of Science for 
financial support. 
The work of G.C.C. is supported in part by the Grant-in-Aid for 
Science Research, Ministry of Education, Science and Culture, 
Japan (No.15740146).

\end{document}